\documentclass[prd,preprint,tightenlines,floatfix,showpacs,preprintnumbers,nofootinbib,eqsecnum,superscriptaddress]{revtex4}

 \usepackage[dvips,final]{graphicx}
  \usepackage{amssymb}
   \usepackage{amsmath}
    \usepackage{amsfonts}
     \usepackage{epsfig}
      \usepackage{bm}% bold math
\begin{document}

\title{
Low-energy pion-pion scattering \\
in the \mbox{\boldmath $pp \to pp\pi^{+}\pi^{-}$ and $p\bar p \to p\bar p\pi^{+}\pi^{-}$} reactions}

\author{P.~Lebiedowicz}
\email{piotr.lebiedowicz@ifj.edu.pl}
\affiliation{Institute of Nuclear Physics PAN, PL-31-342 
Cracow, Poland}
 
\author{A.~Szczurek}
\email{antoni.szczurek@ifj.edu.pl}
\affiliation{University of Rzesz\'ow, PL-35-959 Rzesz\'ow, Poland}
\affiliation{Institute of Nuclear Physics PAN, PL-31-342 Cracow,
Poland} 

\author{R.~Kami\'nski}
\email{robert.kaminski@ifj.edu.pl}
\affiliation{Institute of Nuclear Physics PAN, PL-31-342 
Cracow, Poland} 

\begin{abstract}
We evaluate the contribution of pion-pion rescattering
to the $p p \to p p \pi^+ \pi^-$ and 
$p \bar p \to p \bar p \pi^+ \pi^-$ reactions.
We compare our results with the close-to-threshold experimental data.
The pion-pion rescattering contribution is found there to 
be negligible. The predictions for future experiments
with PANDA detector at High Energy Storage Ring (HESR)
in GSI Darmstadt are presented. It is discussed how
to cut off the dominant Roper resonance and 
double-$\Delta$ excitation mechanisms leading 
to the $p p \pi^+ \pi^-$ channel in final state. 
Differential distributions in momentum transfers, 
transverse momentum, effective two-pions mass, relative azimuthal angle
between pions, and pion rapidities are presented.
\end{abstract}

\pacs{13.75.Cs, 13.75.Lb, 11.80.Et}
%Keywords:

\maketitle

%--------------------------------------------------
\section{Introduction}
%--------------------------------------------------

\begin{figure}[!h]    % Figure 1
\includegraphics[width=0.4\textwidth]{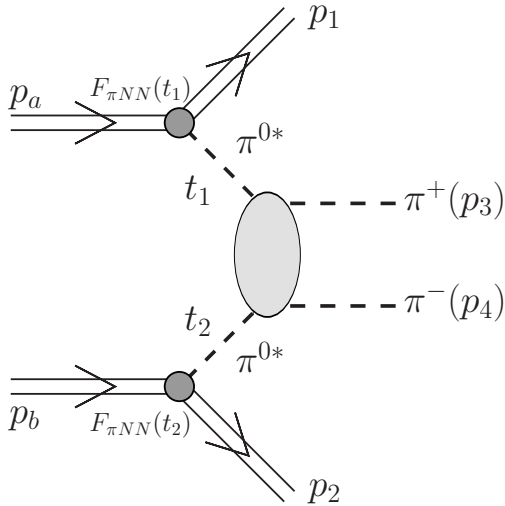}
   \caption{\label{fig:pion-pion-rescattering}
   \small 
The pion-pion rescattering mechanisms of exclusive 
production of $\pi^+$ and $\pi^-$ in proton-proton and 
proton-antiproton collisions. Some kinematical variables
are shown explicitly. The stars attached to $\pi^0$ mesons
denote the fact that they are off-mass-shell.
}
\end{figure}

The $p p \to p p \pi^+ \pi^-$ reaction, which is one
of the reactions with four charged particles
in the final state, can be easily measured. 
Very close to the threshold the excitation of 
the Roper resonance and its subsequent decay as well as 
double-$\Delta$ excitation constitute
the dominant reaction mechanism \cite{AOH98}.
Only energy dependence of the total cross section
was discussed in \cite{AOH98}.
The pion-pion rescattering mechanism shown in 
Fig.\ref{fig:pion-pion-rescattering} was not discussed
so far in the literature.
%On the other hand the low-energy pion-pion scattering 
%was studied in details in the literature.
\vspace{0.6cm}
%--------------------------------------------------------
%--------------------------------------------------------

On the other hand a significant progress in studying 
pion-pion scattering at low-energy has been recently 
achieved due to works based on dispersive analyses
of experimental data 
\cite{A4,CGLNPB01,DescotesGenon:2001tn,PY,KPYII,KPYIII}.
These works were sometimes used together with 
Chiral Perturbation Theory (ChPT) to fix amplitude \cite{A4,CGLNPB01}
and sometimes to test ChPT predictions \cite{PY,KPYII,KPYIII}.
They led to precise determination of the $\pi\pi$ scattering 
amplitudes consistent with analyticity,unitarity and crossing symmetry.
Strong theoretical constraints from forward dispersion relations and
sum rules allowed to test several, sometimes conflicting sets of data \cite{PY}.
Together with twice subtracted dispersion relations (the Roy's equations)
these constraints helped to significantly 
diminish the errors of, for example, sigma pole
position and $S$-wave scattering lengths \cite{PY,KPYII,KPYIII}.
Recent works on once subtracted dispersion relations 
confirmed these results \cite{KRGP}.
Application of Roy's equations in another dispersive 
analysis of experimental data allowed to
eliminate the long standing "up-down" ambiguity 
below 1 GeV in $lI$ 
\footnote{Here $l$ is angular momentum between
pions and $I$ is the total isospin of the pion pair.}
= $S0$ wave \cite{Kaminski:2002pe}.
The simple and model independent amplitudes 
of the $S0$, $P$, $S2$, $D0$, $D2$, $F$ and $G$ 
waves 
presented in series of works \cite{PY,KPYII,KPYIII,KRGP} 
seem to be efficient and easy to use in analyses of $\pi\pi$ interactions.
Amplitudes presented in \cite{KPYIII} have been applied in this analysis to
parameterize the final state interactions $\pi^0\pi^0 \to \pi^+\pi^-$. 

The knowledge from the $\pi \pi \to \pi \pi$ reaction can  
almost directly be used in our $p p \to p p \pi \pi$
reaction
shown in Fig.\ref{fig:pion-pion-rescattering}. 
It is the aim of this paper to evaluate the pion-pion
rescattering contribution for the measured 
close-to-threshold region of the 
$p p \to p p \pi^+ \pi^-$ reaction as well as to make
predictions for the future experiments with the PANDA
detector at High Energy Storage Ring (HESR) in 
GSI Darmstadt for the 
$p \bar p \to p \bar p \pi^+ \pi^-$ reaction.

%----------------------------------------
\section{The two-pion rescattering amplitude}
%----------------------------------------

It is straightforward to evaluate the pion-pion
exchange current contribution shown in 
Fig.\ref{fig:pion-pion-rescattering}.
If we assume the $i \gamma_5$ type coupling of the pion 
to the nucleon then the Born amplitude squared
and averaged over initial and summed over final spin 
polarizations reads:
\begin{eqnarray}
\overline{|{\cal M}|^2}=\dfrac{1}{4}&
\left[ \left( E_a + m \right)\left( E_1+ m \right)
\left(\dfrac{{\bf p}_a^2}{(E_a + m)^2} + \dfrac{{\bf p}_1^2}{(E_1 + m)^2} -
\dfrac{ 2 {\bf p}_a \cdot {\bf p}_1}{(E_a + m)(E_1 + m)}\right)\right] \times 2 \nonumber \\
  \times &
\dfrac{g_{\pi NN}^2}{(t_1 - m_{\pi}^2)^2} F_{\pi NN}^2(t_1)
\; \times \; |{\cal M}_{\pi^{0*}\pi^{0*}\to\pi^+\pi^-}(s_{34},cos\theta^{*}, t_{1},t_{2})|^2 
\; \times \; \dfrac{g_{\pi NN}^2}{(t_2 - m_{\pi}^2)^2} F_{\pi NN}^2(t_2) \nonumber \\
  \times &
\left[ \left( E_b + m \right)\left( E_2+ m \right)
\left(\dfrac{{\bf p}_b^2}{(E_2 + m)^2} + \dfrac{{\bf p}_2^2}{(E_2 + m)^2} -
\dfrac{ 2 {\bf p}_b \cdot {\bf p}_2}{(E_b + m)(E_2 + m)}\right)\right] \times 2 \; . \nonumber \\
\label{pion-pion_amplitude}
\end{eqnarray}
In the formula above $m$ is the mass of the nucleon,
$E_a, E_b$ and $E_1, E_2$ are the energies of initial and outgoing nucleons,
${\bf p}_a, {\bf p}_b$ and ${\bf p}_1, {\bf p}_2$ are the corresponding
three-momenta and $m_{\pi}$ is the pion mass.
The factor $g_{\pi NN}$ is the familiar pion nucleon coupling constant 
and is relatively well known \cite{ELT02} ($\frac{g_{\pi NN}^2}{4 \pi}$ = 13.5 -- 14.6).

%The isospin factor $T_k$ equals 1 for the $\pi^0 \pi^0$ fusion
%and equals 2 for the $\pi^+ \pi^-$ fusion. 
%In the case of proton-antiproton collisions both $p \bar p \pi^{+}\pi^{-}$ and 
%$p \bar p \pi^{0}\pi^{0}$ or $n \bar n \pi^{+}\pi^{-}$ and 
%$n \bar n \pi^{0}\pi^{0}$ final state channel are possible, i.e. both
%$\pi^0 \pi^0$ and $\pi^+ \pi^-$ MEC are allowed. 

In the case of central production of pion pairs not far from the threshold region
rather large transfer momenta squared $t_1$ and $t_2$ are involved
and one has to include non-point-like and off-shellness nature of the particles 
involved in corresponding vertices. This is incorporated via the $F_{\pi NN}(t_1)$
or $F_{\pi NN}(t_2)$ vertex form factors. We shall discuss how the uncertainties 
of the form factors influence our final results. 
In the meson exchange approach \cite{MHE87} they 
are parameterized in the monopole form as
\begin{equation}
F_{\pi N N}(t_{1,2}) = \frac{\Lambda^2 - m_{\pi}^2}{\Lambda^2 - t_{1,2}} \;.
\label{F_piNN_formfactor}
\end{equation} 
In the following for brevity we shall use notation $t_{1,2}$ 
which means $t_1$ or $t_2$.
Typical values of the form factor parameters are $\Lambda$ = 1.2--1.4 GeV \cite{MHE87},
however the Gottfried Sum Rule violation prefers smaller 
$\Lambda \approx$ 0.8 GeV \cite{GSR}.

The amplitude of the subprocess 
$\pi^{0*}\pi^{0*}\rightarrow\pi^+\pi^-$ with virtual initial pions
is written in terms of the amplitude for real intial pions
and correction factors as:
\begin{eqnarray}
{\cal M}_{\pi^{0*}\pi^{0*}\to\pi^+\pi^-}(s_{34},cos\theta^{*}, t_{1},t_{2})&=&
{\cal M}_{\pi^{0*}\pi^{0*}\to\pi^+\pi^-}(s_{34},cos\theta^{*}) F_{\pi^{0*}}(t_{1})F_{\pi^{0*}}(t_{2})\;.
\label{Regge_amplitude}
\end{eqnarray}
The on-shell amplitude can be expanded into partial wave 
amplitudes
$f_l^I(s_{34})$ with angular momentum $l$ and isospin $I$:
\begin{equation}
{\cal M}(s_{34},cos\theta^{*})= 16\pi 
\sum_I \sum_l(2l+1) P_{l}(cos\theta^{*}) 
f_l^I(s_{34}) \; .
\label{pipi_pipi_on-shell-amplitude}
\end{equation}
For a limited range of $M_{\pi\pi} = \sqrt{s_{34}}$ it is enough to
take only a few partial waves.
In our calculation $f_l^I(s_{34})$
can be parameterized in terms of the pion-pion phase shifts 
$\delta_l^I(s_{34})$ and the inelasticities $\eta_l^I(s_{34})$
taken from \cite{KPYIII}
\begin{equation}
f_l^I(s_{34})= \sqrt{\frac{s_{34}}{s_{34}-4m_{\pi}^2}}
\frac{\eta_l^I(s_{34})e^{2i\delta_l^I(s_{34})}-1}{2i} \;.
\label{f_l^I}
\end{equation}

In the formula above $F_{\pi^{0*}}(t_{1,2})$ are extra 
correction factors due to off-shellness of initial pions. 
We use exponential form factors of the type
\begin{equation} 
F_{\pi^{0*}}(t_{1,2})=\exp\left( \frac{t_{1,2}-m_{\pi}^{2}}{\Lambda^{2}_{off}}\right) \;,
\label{off-shell_form_factors}
\end{equation}
i.e. normalized to unity on the pion-mass-shell. 
In general, the parameter $\Lambda_{off}$ is not known 
but in principle could be fitted to the
experimental data providing that our mechanism is 
the dominant mechanism which can be true only in a limited
corner of the phase space.  
From our general experience in hadronic physics we 
expect $\Lambda_{off}\sim$ 1 GeV.

The $cos\theta^{*}$ in Eq.(2.4) requires a seperate discussion.
In the on-shell case the $cos\theta$ can be expressed
in terms of the two-body Mandelstam invariants $\hat t$
and $\hat u$ in two equivalent ways:
\begin{eqnarray}
 cos\theta_{\hat t}= 1+\dfrac{2 {\hat t}}
{s_{34}-4m_{\pi}^2}, \nonumber \\
 cos\theta_{\hat u}=-1-\dfrac{2 {\hat u}}
{s_{34}-4m_{\pi}^2}.
\label{costheta_on-shell}
\end{eqnarray}
This can be generalized to the case of off-shell initial pions as:
\begin{eqnarray}
 cos\theta_{\hat t}^{*}= 1+\dfrac{2 {\hat t}}
{s_{34}-m_{\pi}^2-m_{\pi}^2-t_{1}-t_{2}}, \nonumber \\
 cos\theta_{\hat u}^{*}=-1-\dfrac{2 {\hat u}}
{s_{34}-m_{\pi}^2-m_{\pi}^2-t_{1}-t_{2}}.
\label{costheta_off-shell_1}
\end{eqnarray}
In our case of the $2 \to 4$ reaction
\footnote{$2 \to 4$ reaction denotes a type of the reaction
with two initial and four final particles.}
we have to deal with off-shell initial pions and an analytical
continuation of formula (\ref{costheta_off-shell_1})
is required. In the following we use the most
straightforward prescription:
\begin{equation}
cos\theta^{*} = \dfrac{1}{2}
(cos\theta_{\hat t}^{*}+cos\theta_{\hat u}^{*})= 
\dfrac{{\hat t}-{\hat u}}
{s_{34}-m_{\pi}^2-m_{\pi}^2-t_{1}-t_{2}} \; .
\label{costheta_off-shell_2}
\end{equation}
The formula above reproduces the on-shell formula 
(\ref{costheta_on-shell}) when
$t_1 \to m_{\pi}^2$ and $t_2 \to m_{\pi}^2$,
is symmetric with respect to $\hat t$ and $\hat u$
and fulfils the requirement -1 $< cos\theta^{*} <$ 1.

The differential cross sections for the $2 \to 4$ reaction
are calculated using the general formula
\begin{equation}
d \sigma = \frac{1}{2s} \overline{ |{\cal M}|^2} (2 \pi)^4 
\delta^4 (p_a + p_b - p_1 - p_2 - p_3 - p_4)
\frac{d^3 p_1}{(2 \pi)^3 2 E_1} 
\frac{d^3 p_2}{(2 \pi)^3 2 E_2} 
\frac{d^3 p_3}{(2 \pi)^3 2 E_3}
\frac{d^3 p_4}{(2 \pi)^3 2 E_4} \;. 
\label{dsigma_for_2to4}
\end{equation}
%
%-----------------------------------------------------
\section{Reactions via Roper resonance excitation and its decay}
%-----------------------------------------------------

\begin{figure}[!h]    % Figure 2
\includegraphics[width=0.45\textwidth]{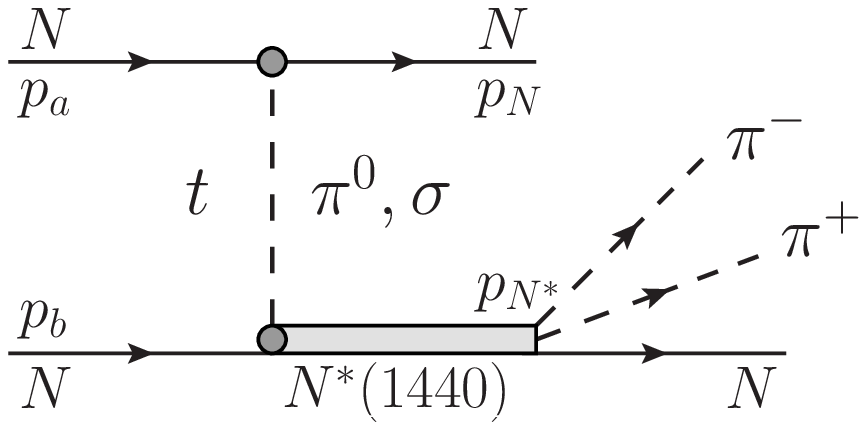}
   \space
\includegraphics[width=0.45\textwidth]{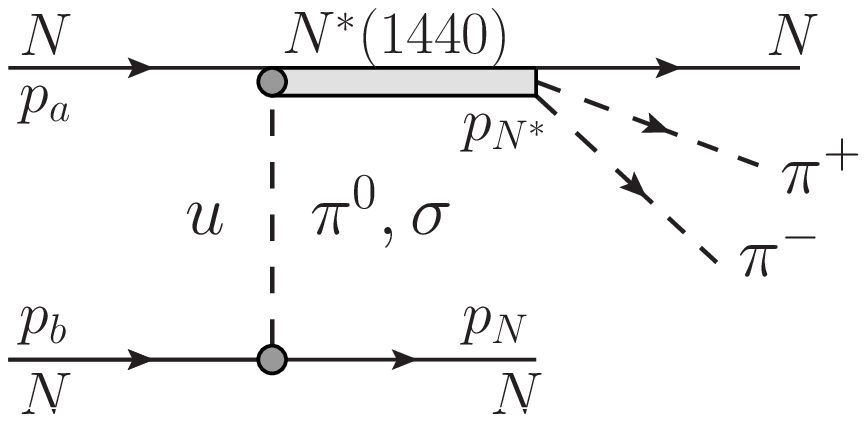}
   \caption{\label{fig:roper_mechanism}
   \small
The dominant mechanisms of Roper resonance production 
at low energy proton-proton scattering.             
}
\end{figure}

Close to the two-pions production threshold the dominant 
mechanism for the reaction $p p \to p p \pi^+ \pi^-$ 
is Roper resonance excitation and its subsequent 
three-body decay \cite{AOH98,OXZX09}. This mechanism constitutes an 
"unwanted background" to our pion-pion rescattering.
At low energy the sigma and pion exchanges are 
the dominant mechanisms of Roper resonance excitation
(see \cite{CLE02}).
Here we show how to approximately estimate the phase-space
integrated contribution of the mechanism shown in 
Fig.\ref{fig:roper_mechanism} not very close to
the threshold \footnote{Very close to the threshold the reaction 
must be treated as genuine four-body reaction with
the $p p \pi^+ \pi^-$ final state (see \cite{AOH98}).}.

%--------------------------------------------------------
%--------------------------------------------------------

The amplitude for the Roper resonance $N^{*}$ excitation via $\sigma$-meson
exchange can be written as
\begin{eqnarray}
{\cal M}_{\lambda_a \lambda_b \to \lambda_N \lambda_{N^*}}^{(\sigma,t)}
&=& g_{\sigma N N} F_{\sigma N N}(t)
[{\bar u}(p_N,\lambda_N)  u(p_a,\lambda_a)]
\frac{1}{t-m_{\sigma}^2}
g_{\sigma N N^*} F_{\sigma N N^*}(t)
[{\bar u}(p_{N^*},\lambda_{N^*}) u(p_b,\lambda_b)]
\;,
\nonumber \\
{\cal M}_{\lambda_a \lambda_b \to \lambda_N \lambda_{N^*}}^{(\sigma,u)}
&=& g_{\sigma N N^{*}} F_{\sigma N N^*}(u)
[{\bar u}(p_{N^*},\lambda_{N^*})  u(p_a,\lambda_a)]
\frac{1}{u-m_{\sigma}^2}
g_{\sigma N N} F_{\sigma N N}(u)
[{\bar u}(p_N,\lambda_N)  u(p_b,\lambda_b)]
\;.\nonumber \\
\label{roper_excitation_sigma_amplitude}
\end{eqnarray}
The amplitude for the Roper resonance excitation via $\pi$-exchange
mechanism can be written as
\begin{eqnarray}
{\cal M}_{\lambda_a \lambda_b \to \lambda_N \lambda_{N^*}}^{(\pi,t)}
&=& g_{\pi N N} F_{\pi N N}(t)
[{\bar u}(p_N,\lambda_N) i \gamma_5 u(p_a,\lambda_a)]
\frac{1}{t-m_{\pi}^2}
\nonumber \\
&&g_{\pi N N^*} F_{\pi N N^*}(t)
[{\bar u}(p_{N^*},\lambda_{N^*}) i \gamma_5 u(p_b,\lambda_b)]\;,
\nonumber \\
{\cal M}_{\lambda_a \lambda_b \to \lambda_N \lambda_{N^*}}^{(\pi,u)}
&=& g_{\pi N N^{*}} F_{\pi N N^{*}}(u)
[{\bar u}(p_{N^*},\lambda_{N^*}) i \gamma_5 u(p_a,\lambda_a)]
\frac{1}{u-m_{\pi}^2}
\nonumber \\
&&g_{\pi N N} F_{\pi N N}(u)
[{\bar u}(p_N,\lambda_N) i \gamma_5 u(p_b,\lambda_b)]\;.
\label{roper_excitation_pion_amplitude}
\end{eqnarray}
In the above equations $g_{\pi N N}$, $g_{\sigma N N}$, $g_{\pi N N^{*}}$, 
$g_{\sigma N N^{*}}$ represent the coupling 
constants and $N$ denotes proton or antiproton;
$u(p_a,\lambda_a)$, $u(p_b,\lambda_b)$,
$u(p_N,\lambda_N)$, $u(p_{N^*},\lambda_{N^*})$, 
are the spionors of the protons and Roper resonance; 
$p_{N}$ and $p_{N^*}$ denote the four-momenta of 
the outgoing proton and the Roper resonance;
$\lambda_{N}$ and $\lambda_{N^*}$ the helicities
of the nucleon and the Roper resonance;
$t$ and $u$ are the four-momentum transfers;
$m_{\pi}$ and $m_{\sigma}$ are the mass of the pion and sigma mesons.

In our calculations the coupling constants are taken as
$g^2_{\pi N N}/4\pi$ = 13.6 \cite{ELT02}, 
$g^2_{\sigma N N}/4\pi$ = 5.69 \cite{MHE87}, 
$g^2_{\pi N N^{*}}/4\pi$ = 2.0 
and $g^2_{\sigma N N^{*}}/4\pi$ = 2.0.
Because numerically the $\sigma$-exchange is the dominant mechanism 
and the $\pi$-exchange is only a small correction 
\footnote{The difference is due to scalar coupling for $\sigma$-exchange or 
pseudoscalar $\gamma_5$ coupling for pion exchange.}, 
in practice the latter can be neglected.
The coupling constant $g_{\sigma N N^{*}}$ is in fact an unknown parameter
which in principle should be determined from the experimental data.
Different values have been used in the literature \cite{MHE87,HCO96}. 
Our number is an average value of those found in the literature.
We parameterize the form factors $F_{\sigma N N}(t,u)$ (and $F_{\pi N N}(t,u)$)
either in the monopole form 
with cut-off parameter $\Lambda_{M}$ as traditionally for low energy processes:
\begin{equation} 
F_{\sigma N N}(t,u) = \frac{\Lambda^{2}_{M} - m_{\sigma}^2}{\Lambda^{2}_{M} - t,u} \;,
\label{monopol_ff}
\end{equation}
or in the exponential form often used at high energies
with cut-off parameter $\Lambda_{E}$:
\begin{equation} 
F_{\sigma N N}(t,u)=\exp\left( \frac{t,u-m_{\sigma}^{2}}{\Lambda^{2}_{E}}\right) \;.
\label{exp_ff}
\end{equation}
The angular distribution for single Roper resonance 
excitation can be calculated from the amplitude above as
\begin{equation}
\frac{d \sigma_{p p \to p N^{*}(1440)}}{d \Omega} =
\frac{1}{64 \pi^2 s} \left( \frac{q_f}{q_i} \right)
\frac{1}{4} 
\sum_{\lambda_a \lambda_b \lambda_N \lambda_{N^*}}
| {\cal M}_{\lambda_a \lambda_b \rightarrow \lambda_N \lambda_{N^*}}^{(a)}(z)
- {\cal M}_{\lambda_a \lambda_b \rightarrow \lambda_N \lambda_{N^*}}^{(b)}(z) |^2,
\label{roper_angular_distribution}
\end{equation}
where $s$ is a square of the proton-proton center-of-mass energy;
$q_{i}$ and $q_{f}$ are center-of-mass momenta in the initial $pp$ or the final 
$pN^{*}$ systems, respectively and $z=cos\theta$, where 
$\theta$ is the center-of-mass angle between the outgoing
and initial nucleon.
The factor $\frac{1}{4}$ and $\sum_{\lambda_a \lambda_b \lambda_N \lambda_{N^*}}$
emerge for the simple reason that the polarization 
of initial and final particles is not considered.
%?
In general, one should calculate the cross section for 2 $\to$ 4 reaction
based on diagrams shown in Fig.\ref{fig:roper_mechanism} with
Roper resonance in the intermediate state (in general off-shell particle).
However, for sufficiently high energies
the total cross section for the $p p \pi^+ \pi^-$ 
final state can be written approximately as a cross 
section for the Roper resonance excitation and 
a probability for the $N^{*}(1440) \to p \pi^+ \pi^-$ decay (on-shell approximation):
\begin{equation}
\sigma_{pp \to pp\pi^+\pi^-}(\sqrt{s}) \approx
\sigma_{pp \to p N^*}(\sqrt{s}) \cdot 
Br(N^{*}(1440) \to p \pi^+ \pi^-) \;.
\label{Roper_to_pppipi}
\end{equation}
This formula will be used to calculate the total 
cross section for the Roper resonance mechanism to show as 
a reference for the discussed above two-pion rescattering 
contribution. The branching ratio to the $p \pi^+ \pi^-$
channel is not very well known and the mechanism of the 
Roper resonance decay can be complicated.
Particle Data Book contains only branching fraction for
all $N \pi \pi$ states. Our decay channel ($p \pi^+ \pi^-$) is only one 
out of three possible ($p \pi^0 \pi^0$, $p \pi^+ \pi^-$, $n \pi^+ \pi^0$). 
We take $Br(N^{*}(1440) \to p \pi^+ \pi^-)$ = 0.1.

%In our calculation, we have not included the $p \bar p$ initial state interaction (ISI) and $p \bar p$ final
%state interaction (FSI) factors. For the energies considered here, $Tp$ > 1 GeV, it is well above
%$pp$ threshold, the role of ISI is basically to reduce the cross section by an overall factor with
%little energy dependence [35, 36], and can be equivalently absorbed into the adjustment of
%form factor parameters. For such energies, only small portion of pn in the final state will                                                                 ̄
%be in relative S-wave and their FSI should not play much important role. Usually, the FSI
%plays significant role only for near-threshold meson production.
%[35] C. Hanhart, Phys. Rept. 397 (2004) 155.
%[36] V. Baru, A. M. Gasparyan, J. Haidenbauer, C. Hanhart, A. E. Kudryavtsev and J. Speth,
%     Phys. Rev. C 67 (2003) 024002
%%%%%%%%%%%%%%%%%%%
%With these effects being very important near threshold, we can not make strong
%statements about the exact role played by the new mechanism but our hope is that we
%can get at least an indication of its relevance.

In the next section we shall show our predictions
for several differential distributions in different
variables.

%-------------------------------------------------

%----------------
\section{Results}
%----------------

\begin{figure}[!h]    % Figure 3
\includegraphics[width=0.5\textwidth]{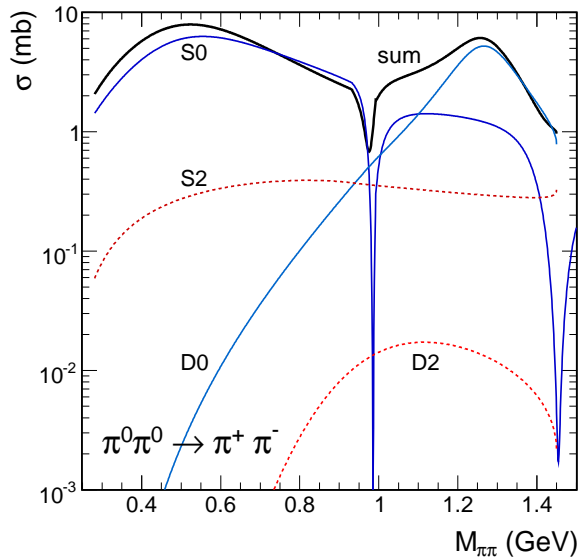}
   \caption{\label{fig:pi0pi0_pippim}
   \small 
The angle-integrated cross section for the reaction 
$\pi^0 \pi^0 \to \pi^+ \pi^-$. The thick solid line represents
the coherent sum of all partial waves. The contributions for
individual partial waves $S0$, $S2$, $D0$ and $D2$ 
are shown for comparison.
}
\end{figure}

Before we go to our four-body reaction let us stay
for a moment with the $\pi^0 \pi^0 \to \pi^+ \pi^-$ on-shell 
scattering.
In Fig.\ref{fig:pi0pi0_pippim} we show the total 
(angle-integrated) cross section for 
the $\pi^0 \pi^ 0 \to \pi^+ \pi^-$ process which
constitutes the subprocess in the $2 \to 4$ reactions
discussed in the present paper. Here the partial wave 
expansion (\ref{pipi_pipi_on-shell-amplitude}) with 
$\delta_l^I$ and $\eta_l^I$ parameterizations from 
Ref.\cite{KPYIII} were used.
In the present work we have limited to 
$M_{\pi \pi} <$ 1.5 GeV, i.e. energies
relevant for WASA and future PANDA experiments.
%--------------------------------------------------------
%--------------------------------------------------------

At higher $\sqrt{s}$ larger $M_{\pi \pi}$ energies 
may contribute. This will be discussed elsewhere 
\cite{LS09}.
We show also individual contributions of different partial
waves: ($l,I$)=(0,0),(0,2),(2,0) and (2,2). 
Because of identity of particles
in the initial state only partial waves with even $l$
contribute. One can see characteristic bumps related
to the famous scalar-isoscalar $\sigma$-meson at 
$M_{\pi \pi} \approx$ 0.5 - 0.6 GeV and the 
tensor-isoscalar $f_2(1270)$. The dip at 
$M_{\pi \pi} =$ 980 MeV is due to interference of the $\sigma$ with another scalar-isoscalar narrow $f_0(980)$ meson. 
Generally the contributions of nonresonant partial waves with $I$=2 
are much smaller than those for $I$=0.

%--------------------------------------------------------
\begin{figure}[!h]    % Figure 4
\includegraphics[width=0.5\textwidth]{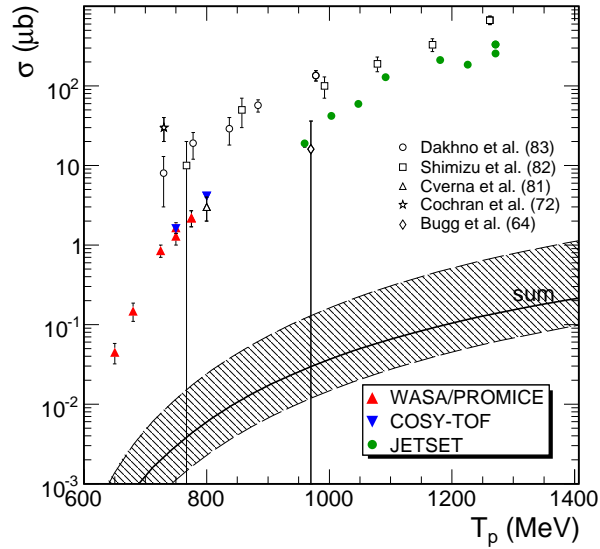}
   \caption{\label{fig:Tp}
   \small 
The phase-space integrated cross section for the $p p \to p p \pi^+ \pi^-$
reaction as a function of the proton kinetic
energy in the laboratory frame $T_p$
together with the experimental data from 
Refs.\cite{BUG64,COC,CVE,SHI,DAK,BUZ,PAT,BRO,JOH,ELB}.
The thick solid line is explained in the text.
The uncertainties band is also shown.
In all cases a coherent sum of all partial waves is taken.}
\end{figure}
%--------------------------------------------------------

In Fig.\ref{fig:Tp} we show the proton energy excitation function 
of the integral cross section for the $p p \to p p \pi^+ \pi^-$ reaction.
In addition, we compare our results with the experimental data
for the $p p \to p p \pi^+ \pi^-$ reaction 
(from Refs.\cite{BUG64,COC,CVE,SHI,DAK,PAT,BRO,JOH,ELB}).
and for the $p \bar p \to p \bar p \pi^+ \pi^-$ one
(only data from the JETSET (PS202) experiment at LEAR \cite{BUZ}).
We present previous data (open symbols) with low statistics 
coming mainly from bubble-chamber measurements
on hydrogen or on deuterium from Refs.\cite{BUG64,COC,SHI,DAK} 
as well as one datum point from an inclusive 
spectrometer measurement at 800 MeV \cite{CVE}.
The newer data taken from Refs.\cite{BUZ,PAT,BRO,JOH,ELB} 
(full symbols) are much closer to the threshold of the reaction
and are an order of magnitude smaller.
%??
%However, the theoretical calculations do not include final state interaction between the protons.
%Including final state interaction in the model would increase the
%predicted total cross section significantly close to the threshold [L. Alvarez-Ruso].
%??
We show how the uncertainties of the form factor 
parameters $\Lambda$ affect our final results. 
For the pion-pion rescattering we modify the cut-off parameter $\Lambda$ 
in Eq.(2.2) $(\Lambda\in$ (0.8, 1.4) GeV) and the cut-off parameter $\Lambda_{off}$ 
in Eq.(2.6) $(\Lambda_{off}\in$ (0.5, 2.0) GeV).
The thick solid line show theoretical predictions from the model
calculations with $\Lambda = 0.8$ GeV and $\Lambda_{off} = 1.0$ GeV.
The pion-pion rescattering contribution is found to be negligible.

\begin{figure}[!h]    % Figure 5
\includegraphics[width=0.5\textwidth]{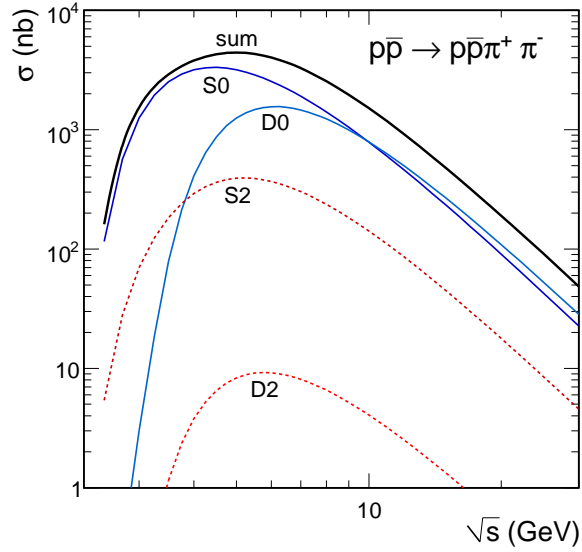}
   \caption{\label{fig:sig_tot_w}
   \small 
The phase-space integrated cross section for the reaction
$p \bar p \to p \bar p \pi^+ \pi^-$ as a function of 
center of mass energy $\sqrt{s}$. The thick solid line represents
the coherent sum of all partial waves. The contributions for
individual partial waves $S0$, $S2$, $D0$ and $D2$ 
are shown for comparison.
}
\end{figure}

As discussed in the theory section, the 
$\pi^0 \pi^0 \to \pi^+ \pi^-$ amplitude used for
the $\pi^0 \pi^0 \to \pi^+ \pi^-$ reaction can, after
a small "correction" for virtualities of both initial 
$\pi^0$'s, be used for the four-body process of our main interest.
In Fig.\ref{fig:sig_tot_w} we show the total cross section
(integrated over the whole phase space with 
the restriction $M_{\pi\pi} <$ 1.42 GeV)
for the four-body reaction as a function of
the overall center-of-mass energy $\sqrt{s}$. We show
the coherent sum of partial waves for different $l$ and $I$
as well as individual contributions. The maximum
of the cross section occurs at $\sqrt{s} \approx$ 5 GeV, 
i.e. at the highest energy planned for the FAIR HESR.
The $l$ = 0, $I$ = 0 partial wave has the dominant contribution.
The sum of the individual contributions is not equal
to the cross section calculated with the sum of the 
partial wave amplitudes because of relatively
strong interference effects.

%--------------------------------------------------------
%--------------------------------------------------------

\begin{figure}[!h]    % Figure 6
\includegraphics[width=0.5\textwidth]{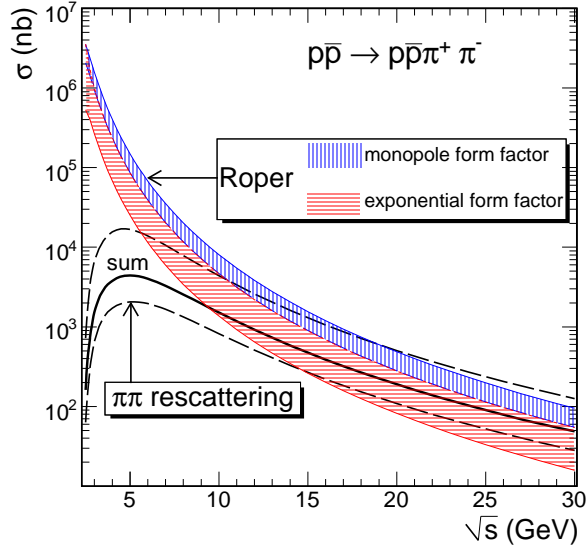}
   \caption{\label{fig:roper_w_minus}
   \small 
The phase-space integrated cross section for the reaction
$p \bar p \to p \bar p \pi^+ \pi^-$ as a function of 
center of mass energy $\sqrt{s}$. We compare the pion-pion 
rescattering contribution and the Roper resonance contribution 
(only $\sigma$-meson exchange included). 
The uncertainty bands for both contributions are also shown.
The area of uncertainties for the pion-pion rescattering contribution is indicated by 
the dashed lines.
%The dashed lines indicates the uncertainties of the pion-pion rescattering contribution.
The pion--pion rescattering contribution is a coherent sum of all partial waves.
}
\end{figure}

In Fig.\ref{fig:roper_w_minus} we compare the pion-pion 
rescattering contribution and the contribution of 
Roper resonance excitation 
through $\sigma$-meson exchange.
In both cases we have estimated the uncertainties of
the contributions. For the pion-pion rescattering 
we modify $\Lambda$ in Eq.(2.2) ($\Lambda \in (0.8, 1.4)$ GeV)
and $\Lambda_{off}$ in Eq.(2.6) ($\Lambda_{off} \in (0.5, 2.0)$ GeV).
The bottom dashed line was obtained with $\Lambda = 0.8$ GeV and $\Lambda_{off} = 0.5$ GeV
while the top dashed line with $\Lambda = 1.4$ GeV and $\Lambda_{off} = 2.0$ GeV.
For the contribution of the Roper resonance excitation through 
$\sigma$-meson exchange we modify 
$\Lambda_{M} \in (1.5, 2.0)$ GeV (band with vertical lines) in the monopole parameterization 
and $\Lambda_{E} \in (1.0, 1.5)$ GeV (band with horizontal lines) in the exponential parameterization.

%--------------------------------------------------------

%--------------------------------------------------------

Because at low energies the Roper resonance excitation and 
double-$\Delta$ excitation play the dominant role
it is not obvious how to extract the pion-pion
rescattering contributions.
To cut off the Roper resonance excitation contribution we eliminate
from the phase space those cases when:\\
$(M_{N^*} - \Delta M_{N^*} < M_{134} < M_{N^*} + \Delta M_{N^*})$ or
$(M_{N^*} - \Delta M_{N^*} < M_{234} < M_{N^*} + \Delta M_{N^*})$.\\
To suppress the double-$\Delta$ excitation we eliminate
from the phase space those cases when:\\
$(M_{\Delta}-\Delta M_{\Delta} < M_{13} < M_{\Delta}+\Delta M_{\Delta}$ and
$ M_{\Delta}-\Delta M_{\Delta} < M_{24} < M_{\Delta}+\Delta M_{\Delta})$ or 
\\
$(M_{\Delta}-\Delta M_{\Delta} < M_{14} < M_{\Delta}+\Delta M_{\Delta}$ and
$ M_{\Delta}-\Delta M_{\Delta} < M_{23} < M_{\Delta}+\Delta M_{\Delta})$.\\
Above $M_{ijk} $ and $M_{ik} $ represent effective mass of the
$p\pi\pi$ and $p \pi$ systems, respectively; 
$\Delta M_{N^*}$ and $\Delta M_{\Delta}$ are cut-off parameters.
We take $\Delta M_{N^*}$ = 0.4 GeV and $\Delta M_{\Delta}$ = 0.2 GeV
which are considerably bigger than the decay widths.

\begin{figure}[!h]    % Figure 7
\includegraphics[width=0.45\textwidth]{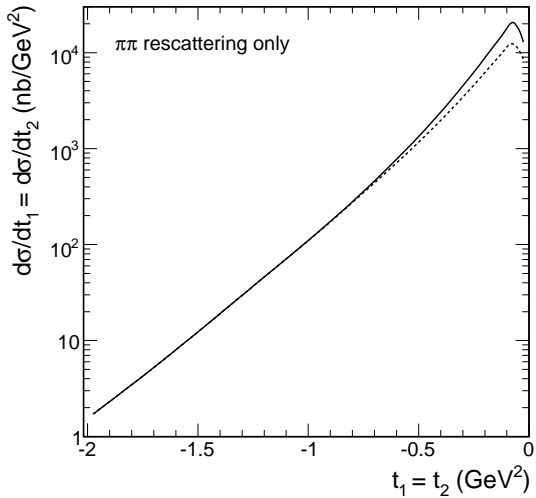}
   \caption{\label{fig:dsig_dt}
   \small 
%Distribution in $t_1$ or $t_2$ for the
Differential cross section 
$\frac{d\sigma}{dt_1} = \frac{d\sigma}{dt_2}$
for the $p \bar p \to p \bar p \pi^+ \pi^-$ reaction
at $\sqrt{s}$ = 5.5 GeV. 
The solid line is the cross section without cuts, the dashed 
line includes cuts to remove regions of Roper 
resonance and double-$\Delta$ excitations.
}
\end{figure}

In Fig.\ref{fig:dsig_dt} we present differential cross section
$\frac{d\sigma}{dt_1} = \frac{d\sigma}{dt_2}$ (integrated
over all other variables) for pion-pion rescattering only. 
The shape of the distribution reflects tensorial structure of 
the $\pi NN$ vertices
(see Eq.(\ref{pion-pion_amplitude})) as well as $t_1$ or $t_2$ dependence
of vertex form factor (see Eq.(\ref{F_piNN_formfactor})).
This plot illustrates how virtual are "initial" pions.
In principle, measuring such distributions would allow
to limit, or even extract, the $\pi NN$ form factor
in relatively broad range of $t_1$ or $t_2$.
This is not possible in elastic nucleon-nucleon 
scattering where many different exchange processes 
contribute.

%--------------------------------------------------------
%--------------------------------------------------------

\begin{figure}[!h]    % Figure 8
\includegraphics[width=0.45\textwidth]{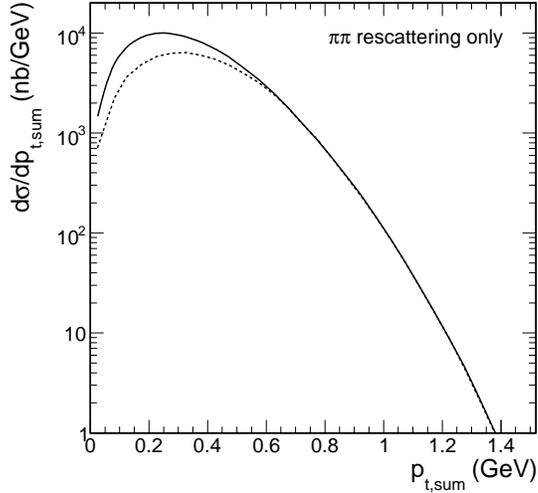}
   \caption{\label{fig:dsig_dptsum}
   \small 
%Distribution in $p_{t,sum}$ 
Differential cross section ${d\sigma}/{dp_{t,sum}}$
for the $p \bar p \to p \bar p \pi^+ \pi^-$ reaction
at $\sqrt{s}$ = 5.5 GeV. 
The solid line is the cross section without cuts, the dashed 
line includes cuts to remove regions of Roper 
resonance and double-$\Delta$ excitations.
}
\end{figure}

In Fig.\ref{fig:dsig_dptsum} we show differential
cross section ${d\sigma}/{dp_{t,sum}}$,
where $\vec{p}_{t,sum} = \vec{p}_{3t}(\pi^+) + \vec{p}_{4t} (\pi^-)$.
For collinear (parallel to the parent nucleons) initial pions
this distribution would be proportional to the Dirac
$\delta(p_{t,sum})$. The deviation from $\delta(p_{t,sum})$
is therefore a measure of noncollinearity and is strongly
related to virtualities of "initial" pions
(see Fig.\ref{fig:dsig_dt}).

%--------------------------------------------------------
%--------------------------------------------------------

\begin{figure}[!h]    % Figure 9
\includegraphics[width=0.45\textwidth]{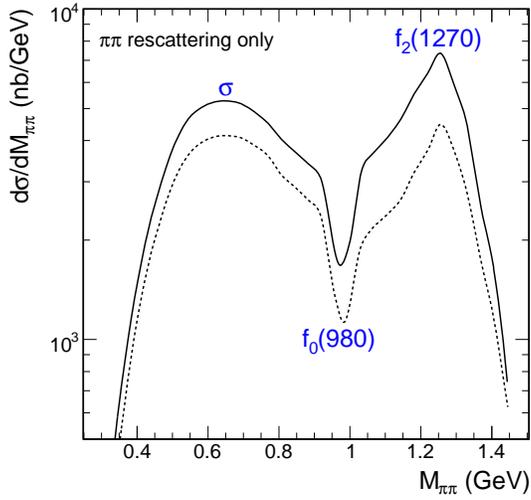}
   \caption{\label{fig:dsig_dmpipi}
   \small 
Differential cross section ${d\sigma}/{dM_{\pi\pi}}$
%Distribution in $\pi^+ \pi^-$ invariant mass 
for the $p \bar p \to p \bar p \pi^+ \pi^-$ reaction
at $\sqrt{s}$ = 5.5 GeV. 
The solid line is the cross section without cuts, the dashed 
line includes cuts to remove regions of Roper 
resonance and double-$\Delta$ excitations.
}
\end{figure}

The two-pion invariant mass distribution given by the differential
cross section ${d\sigma}/{dM_{\pi\pi}}$ is particularly interesting. 
Here (see Fig.\ref{fig:dsig_dmpipi}) one can see two 
characteristic bumps corresponding to the famous scalar-isoscalar
$\sigma $ meson and tensor-isoscalar $f_2(1270)$ meson as well as
the dip from $f_0(980)$.
The cuts to remove regions of Roper resonance 
and double-$\Delta$ excitation only slightly modify 
the spectral shape.

%--------------------------------------------------------
%--------------------------------------------------------

\begin{figure}[!h]    % Figure 10
\includegraphics[width=0.45\textwidth]{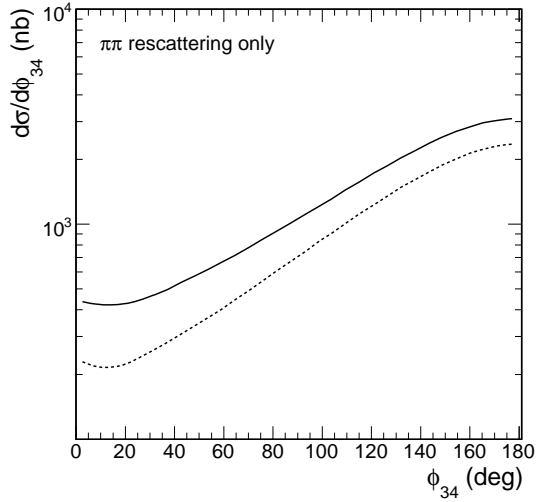}
   \caption{\label{fig:dsig_dphi34}
   \small 
Differential cross section ${d\sigma}/{d\phi_{34}}$
for the $p \bar p \to p \bar p \pi^+ \pi^-$ reaction
at $\sqrt{s}$ = 5.5 GeV. 
The solid line is the cross section without cuts, the dashed 
line includes cuts to remove regions of Roper 
resonance and double-$\Delta$ excitations.
}
\end{figure}

The PANDA detector is supposed to be a $4\pi$ solid angle detector 
with good particle identification for charged particles and photons.
This opens a possibility to study several correlation observables
for outgoing particles. One of them is azimuthal angle correlation
between charged outgoing pions $\phi_{34}$, never discussed in the literature.
In Fig.\ref{fig:dsig_dphi34} we present differential
cross section $d\sigma/d\phi_{34}$.
Clearly a preference of back-to-back emissions can be seen.
Imposing cuts on the Roper resonance and double-$\Delta$ excitation 
lowers the cross section but only mildly modifies the shape.
Because the shape of the azimuthal angle correlations strongly
depends on the reaction mechanism, measuring such correlation
would provide then very valuable information.

%--------------------------------------------------------
%--------------------------------------------------------

\begin{figure}[!h]    % Figure 11
\includegraphics[width=0.45\textwidth]{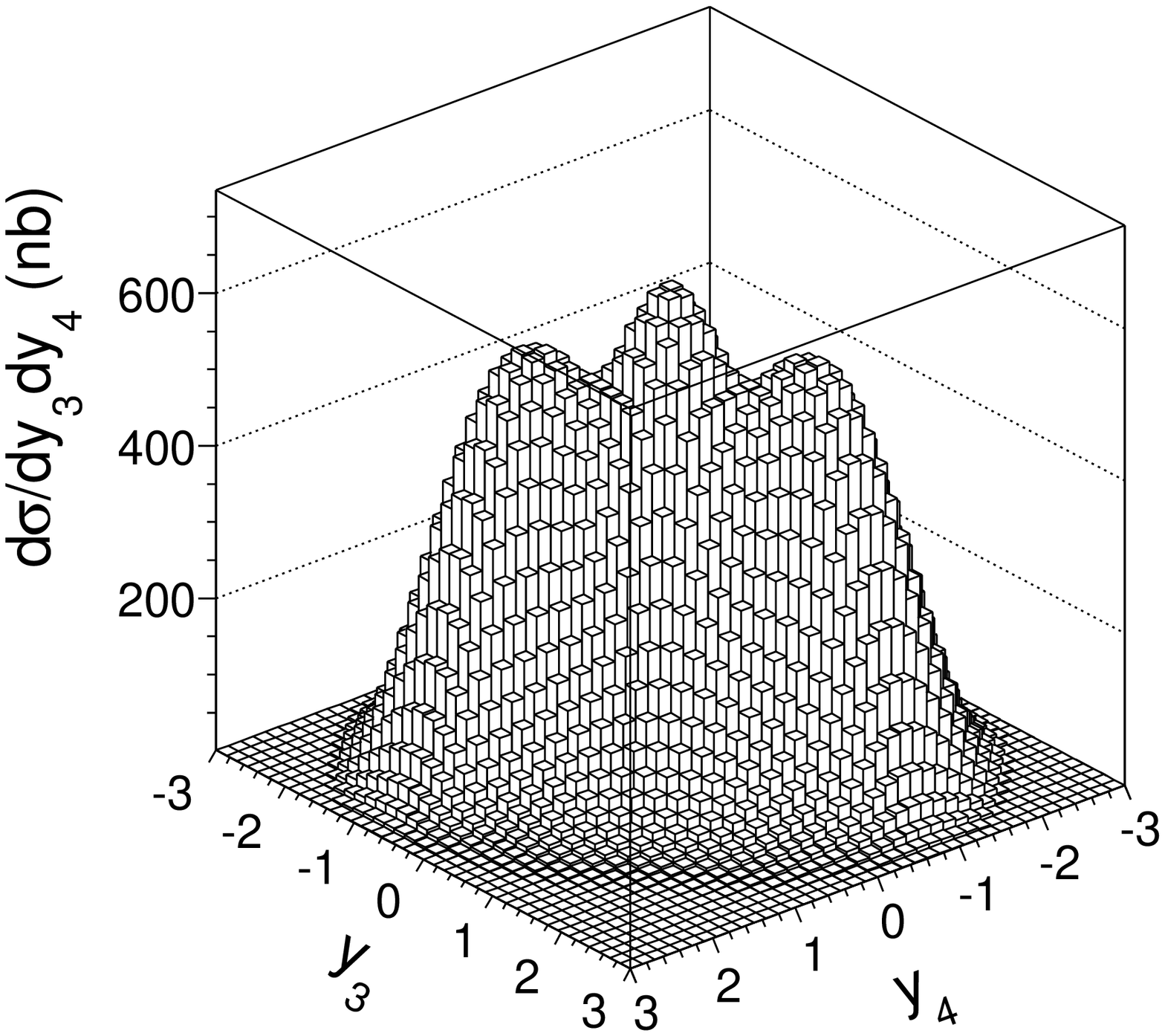}
\includegraphics[width=0.45\textwidth]{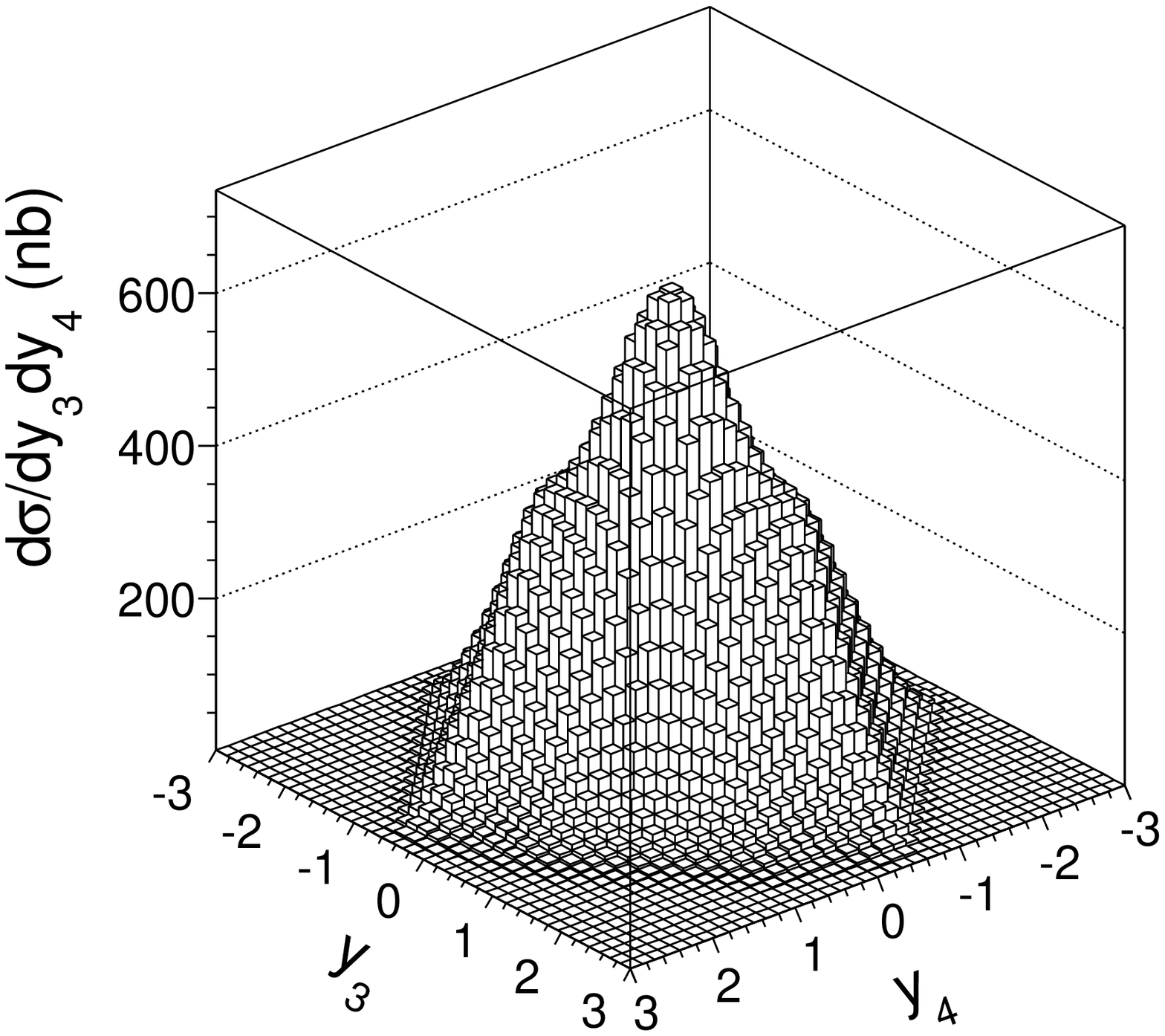}
   \caption{\label{fig:dsig_dy3dy4}
   \small 
Two-dimensional differential cross section $d\sigma/dy_{3}dy_{4}$
in $y_3(\pi^+) \times y_4(\pi^-)$ 
for the $p \bar p \to p \bar p \pi^+ \pi^-$ reaction
at $\sqrt{s}$ = 5.5 GeV (left panel). In the right panel
we have included in addition cuts to remove regions of Roper 
resonance and double-$\Delta$ excitations.
}
\end{figure}

In Fig.\ref{fig:dsig_dy3dy4} we show differential cross section
${d\sigma}/{dy_{3}dy_{4}}$ in the two-dimensional space $(y_3,y_4)$. 
For comparison in the right panel we show a similar distribution
when extra cuts to remove regions of Roper resonance 
and double-$\Delta$ excitation are imposed. 
The cuts do not much affect the region of $y_3, y_4 \approx$ 0.
In practice, the cuts on the Roper resonance
region do not modify the results.
The cuts on Roper resonance act for $(y_{3}<0$ and $y_{4}<0)$
or $(y_{3}>0$ and $y_{4}>0)$ i.e. in the region where
the two-pion rescattering contribution is small. 
The cuts on double-$\Delta$ excitation act for 
$(y_{3}<0$ and $y_{4}>0)$ or $(y_{3}>0$ and $y_{4}<0)$ 
i.e. in the region where the two-pion rescattering 
contribution is sizeable. This shows that the double-$\Delta$
excitation is more critical than the Roper resonance excitation
in the context of extracting the pion-pion rescattering
contribution.

%--------------------------------------------------------
%--------------------------------------------------------

%--------------------
\section{Conclusions}
%--------------------

We have calculated both differential and total cross
sections for the $p p \to p p \pi^+ \pi^-$ and
$p \bar p \to p \bar p \pi^+ \pi^-$ reactions close to threshold
and for future PANDA experiments.
Our results have been
compared with very close to threshold data measured
by the WASA collaboration. We have shown that very
close to threshold the pion-pion rescattering mechanism
gives much smaller contribution than the excitation
of the Roper resonance and its subsequent decay as well as the 
double-$\Delta$ excitation and subsequent decays studied
in the past \cite{AOH98}.
At low energies all these mechanisms overlap and it is
not possible to extract the pion-pion rescattering
contributions and therefore not possible to study
the $\pi^0 \pi^0 \to \pi^+ \pi^-$ process.

Going to higher energies allows to find regions
of the final state phase space where the pion-pion rescattering process
dominates over the Roper resonance and double-$\Delta$ excitation 
mechanisms.
Experiments at highest energies of the HESR (FAIR project)
at GSI Darmstadt open a possibility to study the pion
rescattering process 
and provide an excellent place for studying properties of 
the Roper $N^{*}(1440)$ resonance.

We have presented several distributions which could be
measured in the future with the PANDA detector at the GSI HESR.
Particularly interesting is the distribution
in two-pion invariant mass, where one should observe
bumps related to the famous scalar-isoscalar $\sigma$-meson
and to tensor-isoscalar $f_2(1270)$ meson as well as
a dip from the interference with $f_0(980)$ and $\sigma$.
This distribution is slightly different compared to the
dependence of the total $\pi^0 \pi^0 \to \pi^+ \pi^-$
cross section on $M_{\pi \pi}$.
This is caused mainly by the four-body phase space
modifications.

The pions from the pion-pion rescattering are
produced preferentially in opposite hemispheres,
i.e. if one pion is produced at positive center-of-mass 
rapidities the second pion is produced at negative ones. 
This is similar to the double-$\Delta$ excitation mechanism.
Imposing cuts on double-$\Delta$ excitation leaves
untouched the region of midrapidities.
Also the region of large $p_{t,sum}$ stays unmodified
by the cuts on double-$\Delta$ excitations.

\vspace{1cm}

%--------------------
{\bf Acknowledgments}
%--------------------

Discussions and communications with Benoit Loiseau
are acknowledged. This work was partially supported
by the Polish Ministry of Science and Higher Education (MNiSW) 
under Contract MNiSW N N202 249235.

%---------------------------------------------------------------------

%---------------------------------------------------------------------
%---------------------------------------------------------------------

\end{document}